\documentclass[twocolumn]{aa}
\usepackage{graphicx}
\usepackage{txfonts}
\begin{document}
   \title{Interpreting microlensing signal in QSO~2237$+$0305:\\
         Stars or planets?}


   \author{Rodrigo Gil-Merino and Geraint F. Lewis
          }

   \offprints{R. Gil-Merino}

   \institute{Institute of Astronomy, School of Physics,
              University of Sydney, NSW 2006, Australia\\
              \email{rodrigo, gfl@physics.usyd.edu.au},             
	     }

   \date{Received ; accepted }

   \abstract{  The  multiply  imaged, gravitationally  lensed  quasar,
QSO~2237$+$0305,  has been  the subject  of recent  optical monitoring
campaigns, with  its light curves  displaying uncorrelated variability
attributed to  gravitational microlensing by masses  in the foreground
galaxy.  Based  on these  light curves, it  has been claimed  that the
dominant microlensing population must  be a population of free-floating
Jupiter-like  objects; such  a  conclusion is  not  new, with  several
similar claims  in the literature.  Historically, however, it  has been
shown   that  such   conclusions   are  flawed,   with  an   incorrect
interpretation  of   the  complex  caustic  network   that  arises  at
significant optical depth. This paper examines this more recent claim, 
illustrating that it too is flawed.  

   }

\titlerunning{Microlensing in QSO~2237$+$0305: Stars or planets?}
\authorrunning{Gil-Merino \& Lewis}

\maketitle

\section{Introduction}

The gravitational  lens system  QSO~2237$+$0305 consists of  a distant
quasar, at $z_Q=1.695$, quadruply imaged by a nearby spiral galaxy, at
$z_G=0.039$  (Huchra  et  al.    1985).   Soon  after  its  discovery,
photometric  monitoring of  the individual  images revealed  that they
possessed uncorrelated variability; this  was interpreted as being due
to gravitational microlensing by compact objects in the lensing galaxy
(Irwin  et al.   1989).  This  conclusion was  confirmed  with further
monitoring  of this system,  revealing complex  variability consistent
with  the quasar being  swept with  the high  magnifications associated
with caustics  (Corrigan et  al.   1991, {\O}stensen  et al.   1996,
Vakulkik et  al. 1997, Wo\'zniak et  al.  2000, Alcalde  et al.  2002,
Schmidt  et al. 2002).  Two recent  campaigns have  provided exquisite
light curves for  the four images in this  system; OGLE (Wo\'zniak et 
al.  2000) consists of low temporal sampling over a long time frame, 
while GLITP (Alcalde et al. 2002) was a short, targeted campaign, 
undertake at high temporal sampling.

The datasets from  OGLE and GLITP have been  used extensively to probe
the physical  properties of  both the quasar  emission region  and the
distribution of microlensing masses  in the foreground galaxy.  Wyithe
et al.  (2000a)  used the OGLE (and other extant  datasets) to place a
limit on  the size  of the  quasar emission region,  finding it  to be
smaller than $0.025~\eta_0$ (99\%  confidence level). Here, $\eta_0$ is
the Einstein radius in the  source plane, the natural scale length for
microlensing  (e.g., Schneider et al. 1992);  
this  depends   upon  the  square  root  of  the
microlensing masses, and  for a Solar mass star  in QSO~2237+0305 this
corresponds   to  $1.9\times10^{17}$cm.    For   a  sub-Jupiter   mass
microlensing ($10^{-4}$M$_\odot$) acting as a microlenses, this implies
a source emission radius of $\sim5\times10^{13}$ cm.  In a subsequent
paper,  these  authors  ruled   out  explicitly  and  unambiguously  a
significant   contribution  of  such   (sub-)Jupiter-mass  microlenses
(Wyithe et al 2000b), while Wyithe et al. (2000c) also put limited the
transverse  velocity  of  the  lensing  galaxy  with  an  upper  limit
$v_t<600$ km/s  in the lens plane (which  corresponds approximately to
$v_t<6000$  km/s in  the  source plane), analyzing statistically the
properties  of the  derivatives  of the  amplitude  variations in  the
light curves of the system. Furthermore, Shalyapin et al.  (2002) and 
Goicoechea et al. (2003) analyzed the light curves obtained during the
GLITP observational campaign, utilizing both V and R bands to obtained
a  physical  source  size  of  $3.7\times10^{16}$  cm  (0.012  pc)  at
$2\sigma$  confidence   level.   Moreover,  they   fitted  a  standard
Newtonian  accretion disk for  the source  (Shakura \&  Sunyaev 1973),
amongst others, to  obtain a mass for the central  massive black hole of
$10^8~M_\odot$.



\begin{figure}
   \centering
   \includegraphics[angle=-90,width=6.5cm]{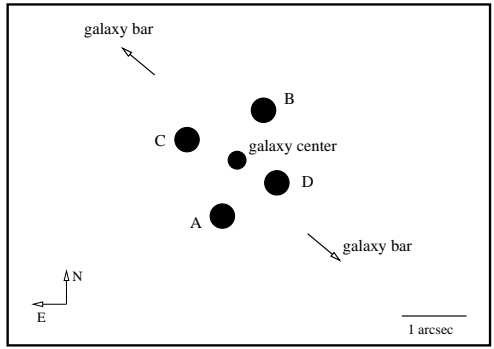}
      \caption{Geometrical configuration of Q2237$+$0305. The four 
               quasar images, the galaxy center and the galaxy bar 
	       are depicted (image motivated by Fig.1 in Witt and 
	       Mao 1994 and taken from Gil-Merino 2003, see also
	       Yee 1988).
              }
         \label{config}
   \end{figure}

In a recent contribution, Lee et  al.  (2005) also used the results of
the  OGLE and  GLITP monitoring  campaigns to  determine  the physical
properties  in QSO~2237$+$0305. Their  study presents  a new  method for
fitting the variability observed in a microlensing light curve, exactly
determining the locations of  the microlensing masses and the emission
profile of the  source, finding that only a source  model with a black
hole can reproduce the features  seen in one of the components (namely
A, see Fig.~\ref{config}).  In addition, they found  that the masses
of  the microlenses  in the  lensing galaxy  must be  of the  order of
$10^{-4}~M_\odot$, and the continuum  source size of the quasar should
be $\approx4\times 10^{15}$ cm.

Clearly, the physical quantities obtained  by Lee et al. (2005) are in
evident disagreement  with these  previous studies of  microlensing in
QSO~2237+0305. In  this contribution, the  approach adopted by  Lee et
al.  (2005) is critically examined  and it is shown that the modeling
technique is based upon flawed assumptions and hence their conclusions
cannot be accepted.

\section{Discussion}

Microlensing  at low optical  depths, where  lensing objects  are well
separated  (except  for cases  of  binary  stars, etc.) can  be  simply
analyzed; an example of this  is microlensing within the Galactic halo
where detailed parameters for  lensing masses and separations, as well
as source profiles,  can be determined from well  sampled light curves
(Paczynski 1986a). Once the density of  stars, and hence the microlensing optical depth,
increases, such a  simple approach is not possible  as the combination
of  the  gravitational  lensing  by  many bodies  results  in  complex
magnification patterns and clustered  caustics (Paczynski  1986b). A 
semi-analytical formulation is still possible for point sources 
(Lewis et al. 1993; Witt 1993), 
but in the case of extended sources, such a situation
rapidly becomes analytically  intractable and numerical approaches are
required. The workhorse for studying gravitational lensing in the high
optical depth  regime has been the inverse  ray-tracing method (Kayser
et  al. 1986,  Wambsganss et  al.  1990).   The  synthetic light curves
derived  from these  simulations depend  upon several  parameters: the
mass of the microlenses, the  distribution of matter in the lens plane
(this is  modeled by  a surface mass  density and the  shear obtained
fitting the observable parameters of  the system to a lens model), the
adopted emission  profile for the  source and the  resulting effective
transverse  velocity  due  to   the  joint  motion  of  source,  lens,
microlenses  and observer.   Even when  representing a  combination of
these parameters, an extraordinary range of potential light curves are
possible   due  to   the  different   random  configurations   of  the
microlensing masses.  A dominant factor, however, is the relative size
of the  source emission  region compared to  the Einstein radius  of a
typical  microlensing  mass; as  the  size  of  the source  region  is
increased, the more the microlensing magnification map is smoothed out,
reducing the degree of variability  in a microlensing light curve.  An
implication  of  this relation  is  that  a  small source  with  small
microlensing masses can induce the  same degree of flux variability as
a large source with correspondingly large masses.

The caustic  network produced at  high optical depth is  very complex,
displaying  regions  of tight,  clustered  caustics, interspaced  with
regions of gently undulating  magnification (see fig. 2).  
Importantly, it has been
shown  (e.g.   Paczynski  1986b)   that  at  such  high  optical  depth
individual microlenses cannot accurately reproduce the complicated net
of caustics seen in magnification maps.  Since the generation
of caustics in a magnification map is a highly non-linear process, the
addition  or  elimination  of  a single  microlens  can  significantly
distort  significant regions of  the  magnification pattern,  typically on  scales
larger  than the  caustic influence  of an  isolated  mass (Wambsganss
1992).   This   was  revealed  during   the  first  analysis   of  the
microlensing light curve of  QSO~2237$+$0305: Irwin et al.  (1989) tried
to  directly estimate  the mass  of  the microlens  responsible for  a
particular event observed  in one of the images  in this system, using
the time scale of the  variability and relating to the Einstein radius
of  an isolated  microlensing mass.   In  a further  analysis of  this
microlensing  light  curve, the  same  group  (Webster  et al.   1991)
suggested that this procedure was not the correct approach, due to the
non-linear  way in  which microlenses  masses combine  to  produce the
caustic structure,  a situation confirmed  in the theoretical  work of
Wambsganss (1992).

It is important to remember  this lesson when examining the
recent study  of microlensing in  QSO~2237$+$0305 by Lee et  al. (2005).
Their  approach  attempts  to  directly and  exactly  reconstruct  the
observed  variability in  the light  curves, by  firstly  adopting the
shear for the two images they analyzed (A and C) from the macromodel of
Witt  et al.   (1995).  They  then  proceed to  construct the  caustic
network by  individually adding  microlenses; the first  mass produces
the simple caustic structure  of an isolated Chang-Refsdal lens (Chang
\& Refsdal 1979),
but subsequent individual masses distort this caustic structure into a more complex
form.   Then,  convolving  this  caustic  map with  a  source  surface
brightness  distribution, the  synthetic light  curve can  be directly
compared  with  the  observed  variability  and the  location  of  the
microlensing  masses and  the source  profile  can be  adjusted via  a
standard minimization  to account  for the observed  variability. From
the  density of microlensing  masses used  to finally  reconstruct the
microlensing light curve, these authors then "deduce" the microlensing
optical depth responsible.

In  critically examining  the approach  by  Lee et  al. (2005),  named
LOHCAM  (Local High Amplification Event Caustic Model),  
a  number  of shortcomings  are    
directly  evident.
The  construction  approach  assumes that  the  microlensing
masses are somehow isolated,  resulting in an isolated caustic network
whose precise properties can  be determined.  While isolated groups of
caustics do occur  at high optical depths, such  groups are not formed
by isolated groups  of stars, rather they occur  due to the non-linear
combination of distributed microlensing masses embedded in a population
as  a  whole.   This   effectively  invalidates  one  of  the  primary
assumptions of  the approach  by Lee et  al. (2005).  This  is further
exacerbated  by  their  treatment   of  the  shear.  Firstly,  Lee  et
al. (2005)  chose the model of  Witt et al.  (1995),  arguing that the
accurate model of  Schmidt et al.  (1998) does not  take account of the
dark  halo of the  lensing galaxy.  However, the  analysis of  Witt et
al. (1995) found the shear at a particular image is strongly dependent
upon the  adopted parameterization of  their mass models and  hence the
selection of a  particular value may not reflect  the true shear.  The
second problem  with their approach  comes in the application  of this
shear; again,  the authors consider  the cluster of  stars responsible
for the microlensing in  QSO~2237$+$0305 to be somehow isolated, subject
only  to the global  shear for  the region.  In reality,  however, the
shear at  any point  in a high  optical depth microlensing  star field
will  differ  significantly from  the  global  shear  due to  coherent
shearing effects  of other masses in  the population. The  only way to
counter  this  is to  assume  that  the  microlensing situation  under
consideration  is  actually  at  low optical  depth  [effectively  the
conclusion  reached by Lee  et al.   (2005)], or  that the  cluster is
somehow truly isolated from  the overall microlensing population. The
first option contradicts all  modeling of the gravitational lensing in
this system, whereas the second is extremely contrived.

   \begin{figure}
   \centering
   \includegraphics[angle=0,width=4cm]{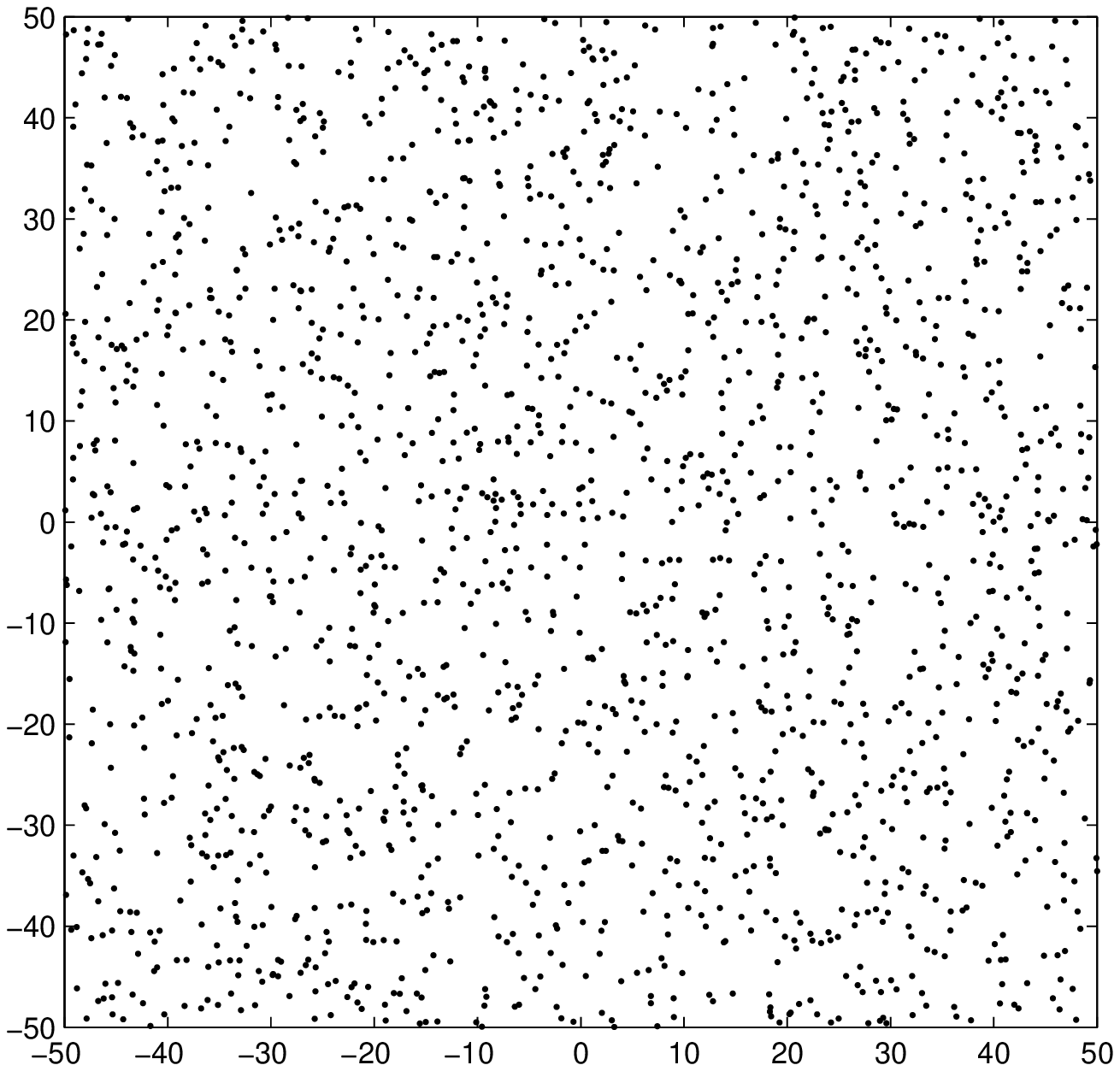}   
   \includegraphics[angle=0,width=6cm]{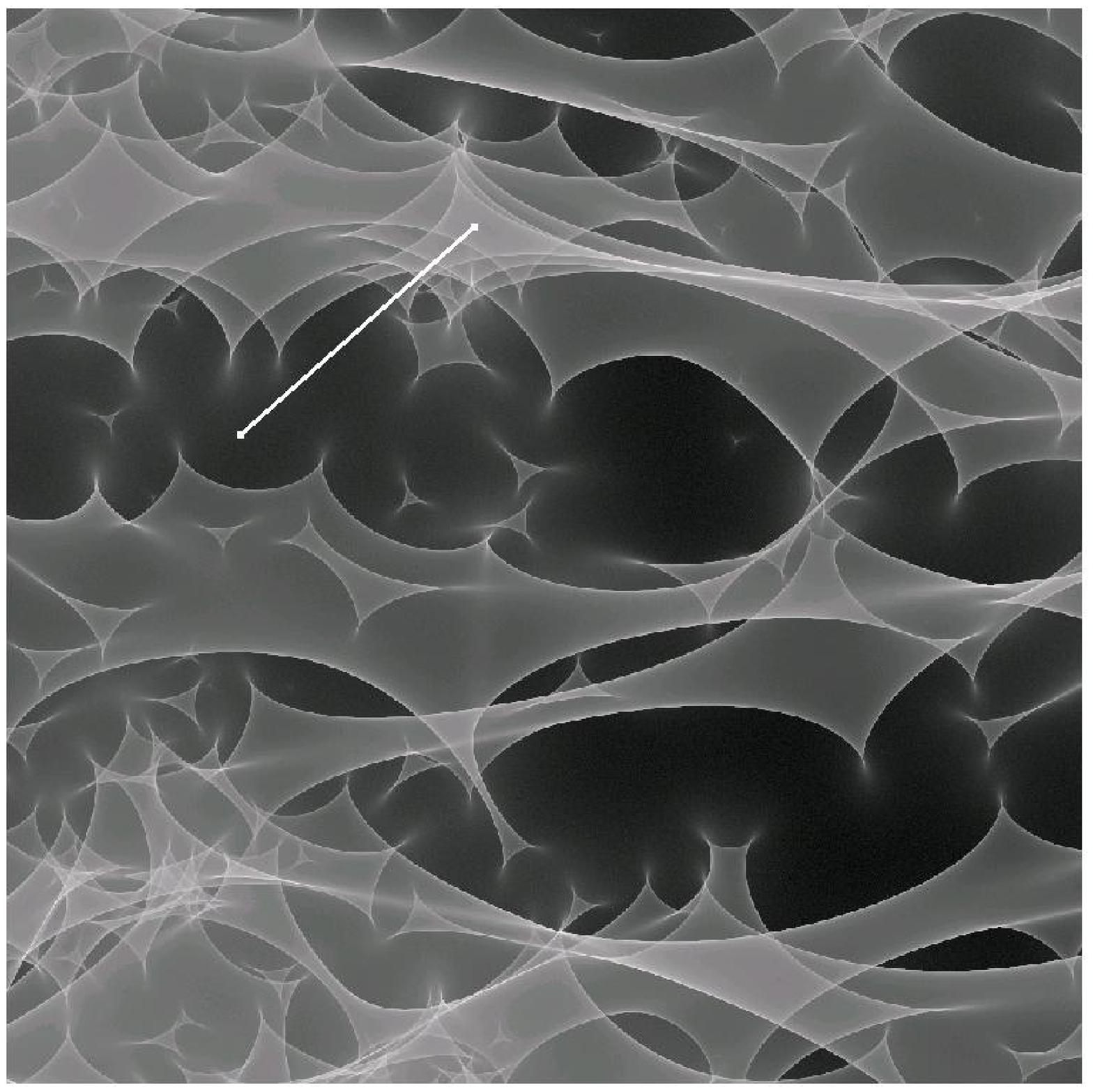}
   \includegraphics[angle=0,width=4cm]{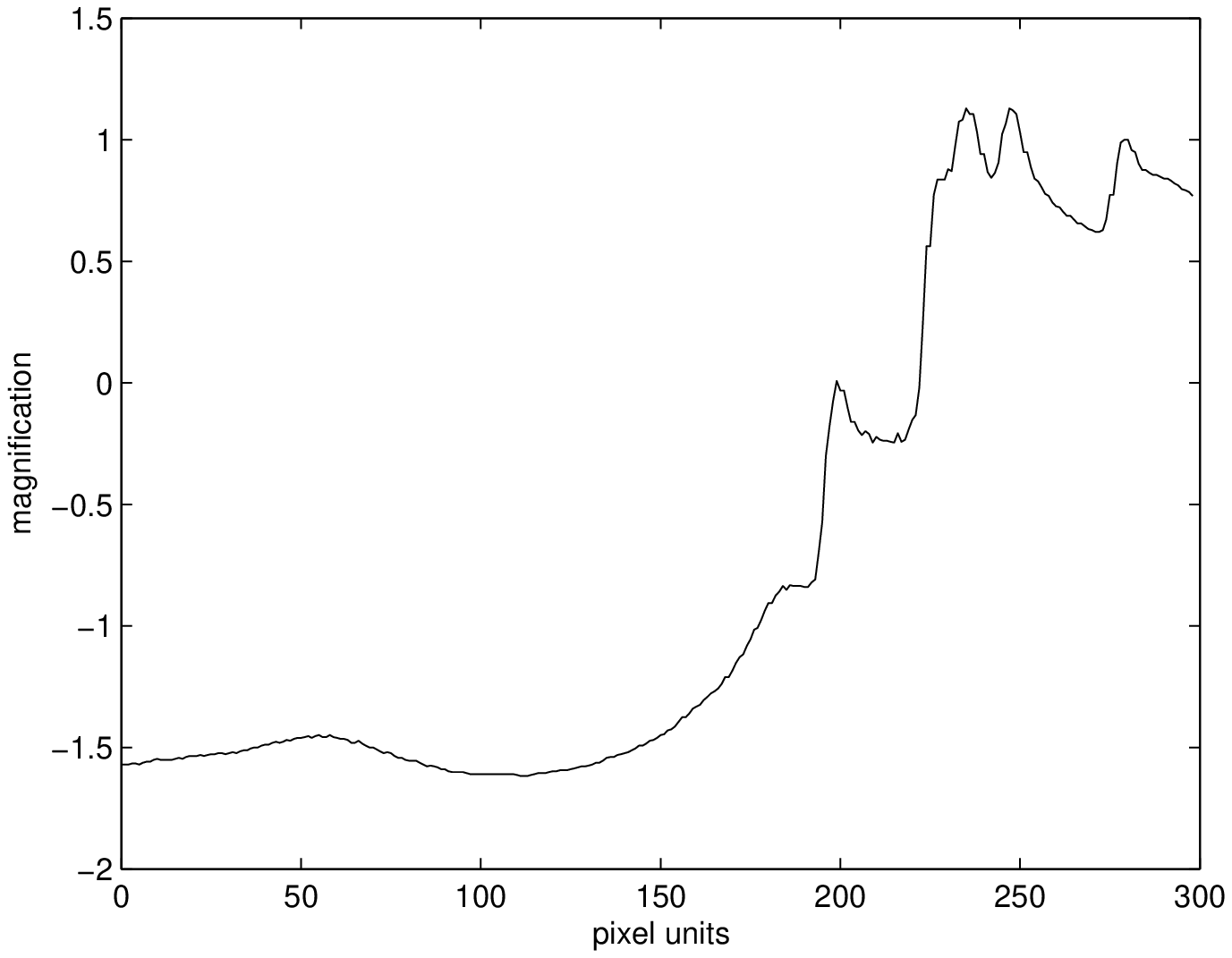}  
      \caption{The upper panel presents a      
       subregion of the random star field for image D of 
      Q2237 using the model
      parameters given by Schmidt el al. (1998), while the central panel
       presents the corresponding 
      caustics pattern is 10 Einstein radii on a side, covered by 
      1024 pixels. This shows regions of very low magnification (dark
      areas) and also complicated caustic clustering (lighter areas).
      Synthetic light curves drawn from such magnification patterns reveal
      both quiescent periods and episodes of higher variability.
      Note that for a typical transverse velocity, the time scale for
      the light curve in this figure (lower panel) is $\sim50$ years, simply accounting  
      for the flat nature of the light curve of image D.
              }
         \label{imageD}
   \end{figure}

The result of this procedure, therefore, will be a macro model for the
images in QSO~2237$+$0305 which bares no resemblance to reality. Rather,
like a typically underconstrained  model, with enough parameters a fit
to the data  can be made, but this fit will  have no physical meaning.
This can clearly be seen if we extend the analysis of Lee et al (2005)
to images B  and D in QSO~2237$+$0305. Unlike images A  and C, these two
images have remained relatively  quiescent since the discovery of this
lensing  system (see Fig.~\ref{config} again for the configuration of
the four images of the system). The  approach  of Lee  et  al. (2005)  would
conclude that  the optical  depth to microlensing  in these  images is
essentially  zero, with no  caustic features  required to  explain the
lack  of  variability. Clearly  this  conflicts  with the  established
models of these images which  show the optical depth to be substantial
and where  quiescent periods naturally arise in  the clustered caustic
maps that result at such optical depths (see fig.~\ref{imageD}). 
It should be noted that such quiescent periods are not due to gaps
in the stellar distribution, but are the result of the complex combination
of magnfications due to many stars in the lensing galaxy.

Hence, the  approach adopted  by  Lee  et al.   (2005)  results in  an
unphysical model, but  could it be employed to  understand the details
of microlensing at high optical  depth? Such an approach would require
no a priori  selection of parameters, such as  global shear, but would
have to have to search  parameter space for particular combinations of
convergence  and  shear.   Such  an approach,  however,  is  extremely
computationally expensive, especially given the almost infinite random
locations of microlensing masses, as  well as the more vexing question
of  just  how  many  masses  are required  to  properly  describe  the
microlensing as  a whole. In summary,  the search for  a unique answer
will  be underconstrained  and  plagued by  degeneracies.  Therefore,  any
realistic   analysis  of   the  microlensing   mass   distribution  in
QSO~2237$+$0305 can only be undertaken via statistical techniques.

Lee et al.  (2005) seem to  be aware of the necessity of a statistical
approach, although  it is not employed in  their current contribution,
and note that it is  necessary to determine the transverse velocity of
the  lensing galaxy in  QSO~2237+0305. Several  statistical approaches
have already been conducted to  determine this velocity (see Wyithe et
al 2000c; Kochanek 2004;  Tuntsov et al. 2004).  Gil-Merino et al. (2005) studied
$1.5\times10^7$ synthetic  light curves drawn  from magnification maps
for images B  and D, examining the distribution  of quiescent periods,
although  all  four images  were  utilized  in  the overall  analysis.
Hence, this analysis is compatible with both periods of quietness for
these  two  components and  periods  of variability  in  the  A and  C
components.  Using  a range of physical source  sizes from $10^{14}$cm
to  $10^{16}$cm,  they  obtained  an  upper limit  for  the  effective
transverse  velocity $v_t<630$  km/s  in the  lens  plane for  $M_{\mu
lenses}=0.1~M_\odot$ (a factor of 10  larger in the source plane and a
factor  of ~3.5  for $M_{\mu  lenses}=1~M_\odot$). The  result  in the
effective transverse velocity was not dependent on the source size in
the  explored range.  Hence,  the source  size adopted  by Lee  et al.
(2005) was indirectly tested in this previous work, which adds another 
argument against the Lee et al. (2005) results; if the conclusions
of Lee et al. (2005) are correct, and a population of sub-Jupiter mass
objects are  responsible for  the microlensing in  QSO~2237$+$0305, then
the transverse velocity of the lensing  galaxy must be reduced by a factor
of  10-100 to be  able to  produce the  quiescent periods  observed in
light curve of image D.  Such a situation is physically implausible as
even if  the bulk velocity of the  galaxy is assumed to  be small, the
peculiar motions of the stars will be much higher (215~km/s, Foltz et
al. 1992) and will be the dominant effect (Kundic \& Wambsganss 1993).

\section{Conclusions}

Lee et al. (2005) have presented a new analysis of the light curves of
the gravitationally microlensed  quasar QSO~2237$+$0305.  The conclusions
of this  result differ significantly  from other analyses of  the same
data set,  finding an exact fit to  the light curves with  only a small
number  of  microlensing  stars.   This letter  has  outlined  several
significant  flaws in  the approach  adopted  by Lee  et al.   (2005),
finding their initial assumptions,  and hence their conclusions, to be
incorrect. Furthermore,  given the  almost infinite potential  range of
configurations of microlensing masses, any  exact fit to the data will
be subject  to a huge  range of degeneracies  and no uniqueness  for a
particular  configuration   can  be  reliably   claimed.  Hence,  such
microlensing analysis  cannot be  represented by a  model minimization
approach. In summary, this letter suggests that the Lee et al. (2005) claim
that their approach will play a key role in the future of the study 
of high magnification microlensing events is incorrect.


\end{document}